\documentclass[preprint,aps,pre,showpacs,superscriptaddress]{revtex4}

\usepackage{amsmath}
\usepackage{amssymb}
\usepackage{graphicx}
\usepackage{bm}

\begin{document}
\title{Lipid membranes with free edges}
\author{Zhan-chun Tu}
\email[Email address: ]{tzc@itp.ac.cn}
\affiliation{Institute of Theoretical Physics,
Academia Sinica, P.O.Box 2735 Beijing 100080,
China} \affiliation{Graduate School, Academia
Sinica, Beijing, China}
\author{Zhong-can Ou-Yang}
\email[Email address: ] {oy@itp.ac.cn}
\affiliation{Institute of Theoretical Physics,
Academia Sinica, P.O.Box 2735 Beijing 100080,
China} \affiliation{Center for Advanced Study,
Tsinghua University, Beijing 100084, China}
\begin{abstract}
Lipid membrane with freely exposed edge is
regarded as smooth surface with curved boundary.
Exterior differential forms are introduced to
describe the surface and the boundary curve. The
total free energy is defined as the sum of
Helfrich's free energy and the surface and line
tension energy. The equilibrium equation and
boundary conditions of the membrane are derived
by taking the variation of the total free energy.
These equations can also be applied to the
membrane with several freely exposed edges.
Analytical and numerical solutions to these
equations are obtained under the axisymmetric
condition. The numerical results can be used to
explain recent experimental results obtained by
Saitoh \emph{et al}. [Proc. Natl. Acad. Sci.
\textbf{95}, 1026 (1998)].
\end{abstract}
 \maketitle
\section{\label{introd}introduction}
Theoretical study on shapes of closed lipid
membranes made great progress two decades ago.
The shape equation of closed membranes was
obtained in 1987 \cite{oy1}, with which the
biconcave discoidal shape of the red cell was
naturally explained \cite{oy2}, and a ratio of
$\sqrt{2}$ of the two radii of a torus vesicle
membrane was predicted \cite{oy3} and confirmed
by experiment \cite{Mutz}.

During the formation process of the cell, either
material will be added to the edge or the edge
will heal itself so as to form closed structure.
There are also metastable cup-like equilibrium
shapes of lipid membranes with free edges
\cite{Boal}. Recently, opening-up process of
liposomal membranes by talin
\cite{Hotani,Hotani2} has also been observed
gives rise to the interest of studying the
equilibrium equation and boundary conditions of
lipid membranes with free exposed edges.
Capovilla \textit{et al.} first study this
problem and give the equilibrium equation and
boundary conditions \cite{Capovilla}. They also
discuss the mechanical meaning of these equations
\cite{Capovilla,Capovilla2}.

The study of these cup-like structures enables us
to understand the assembly process of vesicles.
J\"{u}licher \emph{et al.} suggest that a line
tension can be associated with a domain boundary
between two different phases of an inhomogeneous
vesicle and leads to the budding \cite{Lipowsky}.
For simplicity, however, we will restrict our
discussion on open homogenous vesicles.

In this paper, a lipid membrane with freely
exposed edge is regarded as a differentiable
surface with a boundary curve. Exterior
differential forms are introduced to describe the
surface and the curve. The total free energy is
defined as the sum of Helfrich's free energy and
the surface and line tension energy. The
equilibrium equation and the boundary conditions
of the membrane are derived from the variation of
the total free energy. These equations can also
be applied to the membrane with several freely
exposed edges. This is another way to obtain the
results of Capovilla \textit{et al.} Some
solutions to the equations are obtained and the
corresponding shapes are shown. They can be used
to explain some known experimental results
\cite{Hotani}.

This paper is organized as follows: In
Sec.\ref{extdf}, we retrospect briefly the
surface theory expressed by exterior differential
forms. In Sec.\ref{hodge}, we introduce some
basic properties of Hodge star $\ast$. In
Sec.\ref{variation}, we construct the variational
theory of the surface and give some useful
formulas. In Sec.\ref{functional}, we derive the
equilibrium equation and boundary conditions of
the membrane from the variation of the total free
energy. In Sec.\ref{shape}, we suggest some
special solutions to the equations and show their
corresponding shapes. In Sec.\ref{numeric}, we
put forward a numerical scheme to give some
axisymmetric solutions as well as their
corresponding shapes to explain some experimental
results. In Sec.\ref{conclusion}, we give a brief
conclusion and prospect the challenging work.
\section{\label{extdf}surface theory expressed by exterior differential
forms}
In this section, we retrospect briefly the
surface theory expressed by exterior differential
forms. The details can be found in
Ref.\cite{chen}.

\begin{figure}[htp!]
\includegraphics[width=7cm]{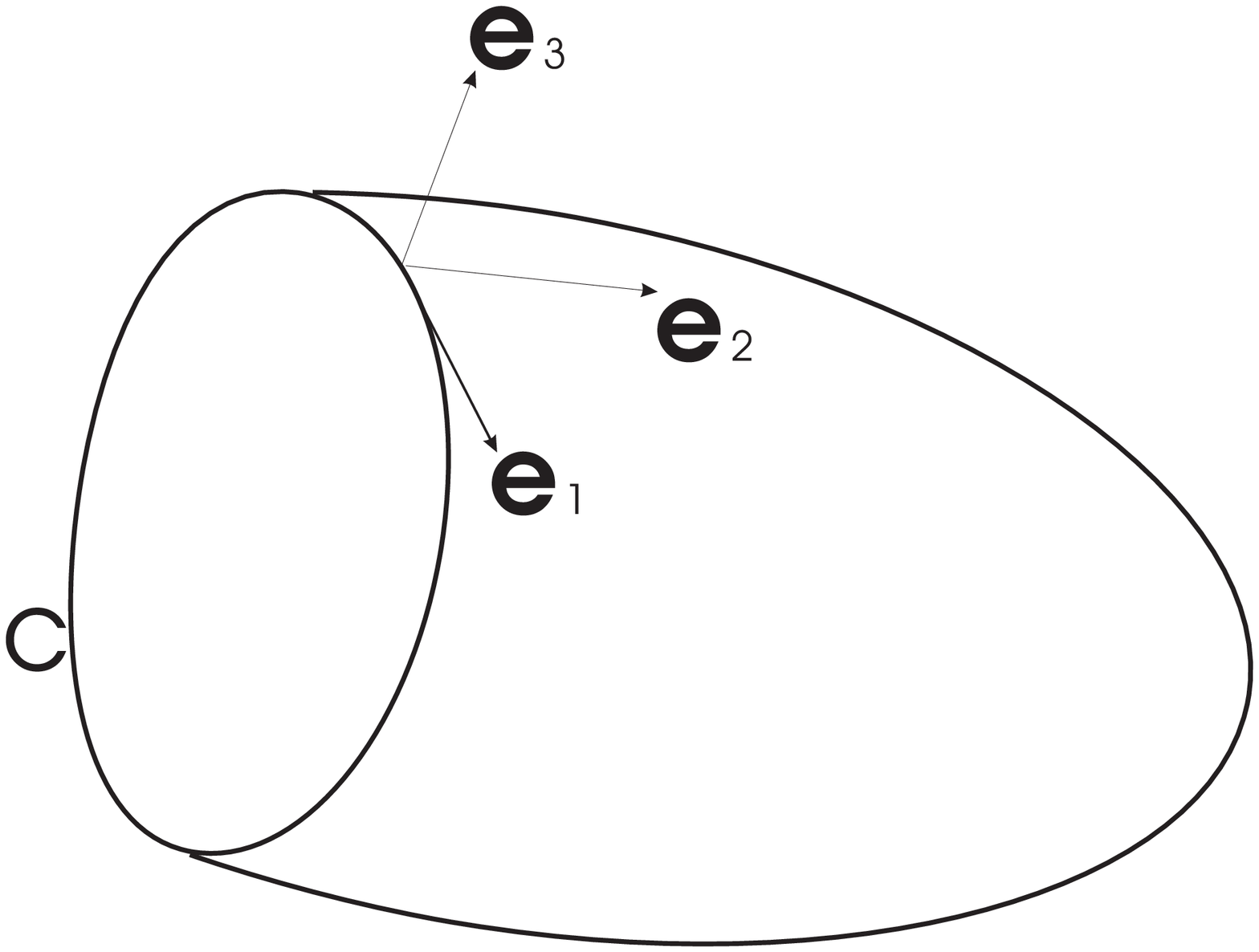}
\caption{\label{fig1}The surface with an edge
$C$. At every point of the surface, we can
construct an orthogonal frame
$\mathbf{e}_1,\mathbf{e}_2,\mathbf{e}_3$, where
$\mathbf{e}_3$ is the normal vector of the
surface. For a point on curve $C$, $\textbf{e}_1$
is the tangent vector of $C$.}
\end{figure}

 We regard a membrane
with freely exposed edge as a differentiable and
orientational surface with a boundary curve $C$,
as shown in Fig.\ref{fig1}. At every point on the
surface, we can choose an orthogonal frame
$\mathbf{e}_1,\mathbf{e}_2,\mathbf{e}_3$ with
$\mathbf{e}_i\cdot\mathbf{e}_j=\delta_{ij}$ and
$\mathbf{e}_3$ being the normal vector. For a
point in curve $C$, $\textbf{e}_1$ is the tangent
vector of $C$.

An infinitesimal tangent vector of the surface is defined as
\begin{equation}
d\mathbf{r}=\omega_1\mathbf{e}_1+\omega_2\mathbf{e}_2,
\end{equation}
where $d$ is an exterior differential operator,
and $\omega_1,\omega_2$ are 1-differential forms.
Moreover, we define
\begin{equation}
d\mathbf{e}_i=\omega_{ij}\mathbf{e}_j,
\end{equation}
where $\omega_{ij}$ satisfies $\omega_{ij}=-\omega_{ji}$ because
of $\mathbf{e}_i\cdot\mathbf{e}_j=\delta_{ij}$.

With $dd=0$ and
$d(\omega_1\wedge\omega_2)=d\omega_1\wedge\omega_2-
\omega_1\wedge d\omega_2$, we have
\begin{equation}  \label{domega1}
d\omega_1=\omega_{12}\wedge\omega_2;\quad
d\omega_2=\omega_{21}\wedge\omega_1;\quad
\omega_1\wedge\omega_{13}+\omega_2\wedge\omega_{23}=0;
\end{equation}
and
\begin{equation}
d\omega_{ij}=\omega_{ik}\wedge\omega_{kj}\quad (i,j=1,2,3),
\end{equation}
where the symbol ``$\wedge$" represents the exterior product on
which the most excellent expatiation may be the Ref.\cite{arnold}.

Eq.(\ref{domega1}) and Cartan lemma imply that:
\begin{equation}  \label{omega13}
\omega_{13}=a\omega_{1}+b\omega_{2},\quad
\omega_{23}=b\omega_{1}+c\omega_{2}.
\end{equation}

Therefore, we have \begin{eqnarray}&&
\texttt{area element: }
dA=\omega_1\wedge\omega_2,\\
&&\texttt{first fundamental form:
}I=d\mathbf{r}\cdot
d\mathbf{r}=\omega_1^2+\omega_2^2,\\
&&\texttt{second fundamental form:
}II=-d\mathbf{r}\cdot
d\mathbf{e}_3=a\omega_1^2+2b\omega_1 \omega_2+c\omega_2^2,\\
&&\texttt{mean curvature: }H=\frac{a+c}{2},\\
&&\texttt{Gaussian curvature: }K=ac-b^2.
\end{eqnarray}

\section{\label{hodge}Hodge star $\ast$}

In this part, we briefly introduce basic properties rather than
the exact definition of Hodge star $\ast$ \cite{westenholz}
because we just use these properties in the following contents.

If $g,h$ are functions defined on 2D smooth surface $M$, then the
following formulas are valid:
\begin{equation}
\ast f=f\omega_1\wedge\omega_2;
\end{equation}
\begin{equation}\label{stardf}
\ast df=-f_2\omega_1+f_1\omega_2, \quad if \quad
df=f_1\omega_1+f_2\omega_2;
\end{equation}
\begin{equation}\label{lap}
d\ast df=\nabla^2 f, \quad \nabla^2\texttt{ is the
Laplace-Beltrami operator}.
\end{equation}

We can easily prove that
\begin{equation}\label{stoke2}
\iint_M(fd\ast dg-gd\ast df)=\oint_{\partial M}(f\ast dg-g\ast df)
\end{equation} through
Stokes's theorem and integration by parts.

\section{\label{variation}variational theory of the surface}

the variation of the surface is defined as:
\begin{equation}
\delta \mathbf{r}=\Omega _{2}\mathbf{e}_{2}+\Omega
_{3}\mathbf{e}_{3},
\end{equation}
where the variation along $\mathbf{e}_{1}$ is
unnecessary because it gives only an identity.
Furthermore, let
\begin{equation}
\delta \mathbf{e}_{i}=\Omega
_{ij}\mathbf{e}_{j},\quad \Omega _{ij}=-\Omega
_{ji}.
\end{equation}

Operators $d$ and $\delta$ are independent, thus
$d\delta =\delta d$. $d\delta \mathbf{r}=\delta
d\mathbf{r}$ implies that:
\begin{eqnarray}
\delta \omega _{1} &=&\Omega _{2}\omega _{21}+\Omega _{3}\omega _{31}-\omega
_{2}\Omega _{21},  \label{detaomega1} \\
\delta \omega _{2} &=&d\Omega _{2}+\Omega _{3}\omega _{32}-\omega _{1}\Omega
_{12},  \label{detaomega2} \\
d\Omega _{3} &=&\Omega _{13}\omega _{1}+\Omega
_{23}\omega _{2}-\Omega _{2}\omega _{23}.
\label{domega3}
\end{eqnarray}
Furthermore, $d\delta \mathbf{e}_i=\delta
d\mathbf{e}_i$ implies that:
\begin{equation}
\delta \omega _{ij}=d\Omega _{ij}+\Omega _{ik}\omega _{kj}-\omega
_{ik}\Omega _{kj}.  \label{detaomegaij}
\end{equation}

It is necessary to point out that the properties
of the operator $\delta$ are exactly similar to
those of the ordinary differential.

\section{\label{functional}equilibrium equation of the membrane and boundary
conditions}

The total free energy $F$ of a membrane with an
edge is defined as the sum of Helfrich's free
energy\cite{helfrich,oy4} $F_H=\iint[
\frac{k_{c}}{2}(2H+c_{0})^{2}+\bar{k}K]dA$ and
the surface and line tension energy
$F_{sl}=\lambda \iint dA+\gamma \oint_{C}ds$.
Here $k_c$, $\bar{k}$, $c_0$, $\lambda$ and
$\gamma$ are constants. With the arc-length
parameter $ds=\omega_1$, the geodesic curvature
$k_g=\omega_{12}/ds$ on $C$ and the Gauss-Bonnet
formula $\iint K dA=2\pi-\oint_C k_g ds$, the
total free energy and its variation are given
\begin{equation}
F=\iint
[\frac{k_{c}}{2}(2H+c_{0})^{2}+\lambda]\omega
_{1}\wedge \omega _{2}+\gamma \oint_{C}\omega
_{1}-\bar{k}\oint_{C}\omega _{12}+2\pi\bar{k},
\end{equation}
and
\begin{eqnarray}
\delta F &=&k_{c}\iint (2H+c_{0})\delta
(2H)\omega _{1}\wedge \omega _{2}+\iint
[\frac{k_{c}}{2}(2H+c_{0})^{2}+\lambda]\delta
(\omega _{1}\wedge \omega
_{2})\nonumber\\&+&\gamma \oint_{C}\delta \omega
_{1}-\bar{k}\oint_{C}\delta \omega
_{12},\label{detaf}
\end{eqnarray}
respectively. From Eqs.(\ref{detaomega1}) and
(\ref{detaomega2}), we can easily obtain:
\begin{eqnarray}
\delta (\omega _{1}\wedge \omega _{2}) &=&\delta \omega _{1}\wedge
\omega _{2}+\omega _{1}\wedge \delta \omega _{2}=-d(\Omega
_{2}\omega _{1})-(2H)\Omega _{3}\omega _{1}\wedge \omega _{2}.
\end{eqnarray}
Eqs.(\ref{omega13}), (\ref{detaomega1}),
(\ref{detaomega2}) and (\ref{detaomegaij}) lead
to:
\begin{eqnarray}
\delta (2H)\omega _{1}\wedge \omega _{2} &=&\delta (a+c)\omega _{1}\wedge
\omega _{2}\nonumber \\&=&2(2H^{2}-K)\Omega _{3}\omega _{1}\wedge \omega _{2}+d(\Omega _{13}\omega _{2}-\Omega _{23}\omega _{1})\nonumber \\
&+&a\Omega _{2}d\omega _{1}-bd\Omega _{2}\wedge \omega
_{2}+b\Omega _{2}d\omega _{2}+cd\Omega _{2}\wedge \omega _{1}.
\end{eqnarray}
Thus we have:
\begin{eqnarray}
\delta F &=&k_{c}\iint (2H+c_{0})[2(2H^{2}-K)\Omega _{3}\omega
_{1}\wedge \omega _{2}+d(\Omega _{13}\omega
_{2}-\Omega _{23}\omega _{1})  \notag \\
&&+a\Omega _{2}d\omega _{1}-bd\Omega _{2}\wedge \omega
_{2}+b\Omega
_{2}d\omega _{2}+cd\Omega _{2}\wedge \omega _{1}]  \notag \\
&&+\iint (\frac{k_{c}}{2}(2H+c_{0})^{2}+\lambda )[-d(\Omega _{2}\omega
_{1})-(2H)\Omega _{3}\omega _{1}\wedge \omega _{2}]  \notag \\
&&+\gamma \oint_{C}[\Omega _{2}\omega
_{21}+\Omega _{3}\omega _{31}-\omega _{2}\Omega
_{21}]-\bar{k}\oint_{C}[d\Omega _{12}+\Omega
_{13}\omega _{32}-\omega _{13}\Omega _{32}].
\label{detaf2}
\end{eqnarray}

If $\Omega _{2}=0$, then $d\Omega _{3}=\Omega
_{13}\omega _{1}+\Omega _{23}\omega _{2}$, $\ast
d\Omega _{3}=-\Omega _{23}\omega _{1}+\Omega
_{13}\omega _{2}$. On curve $C$, $\omega _{2}=0$,
$\omega _{31}=-a\omega _{1}$, $\omega
_{32}=-b\omega _{1}$, $\Omega _{3}|_{C}=\Omega
_{3C}$. Thus Eq.(\ref{detaf2}) is reduced to
\begin{eqnarray}
\delta F &=&\iint [k_{c}(2H+c_{0})(2H^{2}-c_{0}H-2K)-2\lambda
H]\Omega _{3}\omega
_{1}\wedge \omega _{2}\notag \\
&+&k_{c}\iint (2H+c_{0})d\ast d\Omega _3-\gamma \oint_{C}a\omega
_{1}\Omega _{3}+\bar{k}\oint_{C}(b\Omega _{13}-a\Omega
_{23})\omega _{1}
\end{eqnarray}
In terms of Eqs.(\ref{lap}) and (\ref{stoke2}),
we have:
\begin{eqnarray*}
\iint (2H+c_{0})d\ast d\Omega _3=\oint_{C}(2H+c_{0})\ast d\Omega
_3-\oint_{C}\Omega _{3}\ast d(2H+c_{0})+\iint \Omega _{3}\nabla
^{2}(2H+c_{0})\omega _{1}\wedge \omega _{2}.
\end{eqnarray*}
Using integration by parts and Stokes's theorem,
we arrive at $\oint_{C}b\Omega _{13}\omega
_{1}=\oint_{C}bd\Omega _{3C}=-\oint_{C}\Omega
_{3C}db$. Thus
\begin{eqnarray}
\delta F &=&\iint [k_{c}(2H+c_{0})(2H^{2}-c_{0}H-2K)-2\lambda
H+k_{c}\nabla
^{2}(2H+c_{0})]\Omega _{3}\omega _{1}\wedge \omega _{2}\notag \\
&&-\oint_{C}[k_{c}(2H+c_{0})+\bar{k}a]\Omega
_{23}\omega _{1}-\oint_{C}\Omega _{3C}[k_{c}\ast
d(2H+c_{0})+\gamma a\omega _{1}+\bar{k} db].
\end{eqnarray}
It follows that
\begin{eqnarray}
&&k_{c}(2H+c_{0})(2H^{2}-c_{0}H-2K)-2\lambda H+k_{c}\nabla
^{2}(2H) =0,
\label{equlib}\\
&&\left. \lbrack k_{c}(2H+c_{0})+\bar{k}a]\right\vert _{C} =0,\label{bound11} \\
&&\left. \lbrack k_{c}\ast d(2H)+\gamma a\omega _{1}+\bar{k}
db]\right\vert _{C} =0.\label{bound22}
\end{eqnarray}

The mechanical meanings of the above three
equations are: Eq.(\ref{equlib}) is the
equilibrium equation of the membrane;
Eq.(\ref{bound11}) is the moment equilibrium
equation of points on $C$ around the axis
$\mathbf{e}_1$; and Eq.(\ref{bound22}) is the
force equilibrium equation of points on $C$ along
the direction of $\mathbf{e}_3$
\cite{Capovilla,Capovilla2}. It is not surprising
that Eq.(\ref{bound11}) contains the factor
$\bar{k}$ because it is related to the bend
energy in Helfrich's free energy. However, it is
difficult to understand why $\bar{k}$ is also
included in Eq.(\ref{bound22}). In fact, the term
$\bar{k} db$ in Eq.(\ref{bound22}) represents the
shear stress which also contributes to the bend
energy in Helfrich's free energy.

In fact, $a=k_n$ and $b=\tau_g$ are the normal
curvature and the geodesic torsion of curve $C$,
respectively, and $\ast
d(2H)=-\mathbf{e}_2\cdot\nabla(2H)\omega_1$. Thus
Eqs.(\ref{bound11}) and (\ref{bound22}) become
\begin{eqnarray}
&&\left. \lbrack k_{c}(2H+c_{0})+\bar{k}k_n]\right\vert _{C} =0,\label{bound1} \\
&&\left. \lbrack
-k_{c}\mathbf{e}_2\cdot\nabla(2H)+\gamma
k_n+\bar{k} \frac{d\tau_g}{ds}]\right\vert _{C}
=0,\label{bound2}
\end{eqnarray}
respectively.

If $\Omega _{3}=0$, then $d\Omega _{3}=\Omega _{13}\omega
_{1}+\Omega _{23}\omega _{2}-\Omega _{2}\omega
_{23}=(\Omega _{13}-b\Omega _{2})\omega _{1}+(\Omega
_{23}-c\Omega _{2})\omega _{2}=0$. It leads to $\Omega _{13}=b\Omega _{2}$ and $%
\Omega _{23}=c\Omega _{2}$.
\begin{eqnarray}
\delta F &=&k_{c}\iint (2H+c_{0})[a\Omega _{2}d\omega _{1}-bd\Omega
_{2}\wedge \omega _{2}+b\Omega _{2}d\omega _{2}+cd\Omega _{2}\wedge \omega
_{1}+d(\Omega _{13}\omega _{2}-\Omega _{23}\omega _{1})]  \notag \\
&&+\iint [\frac{k_{c}}{2}(2H+c_{0})^{2}+\lambda ][-d(\Omega
_{2}\omega _{1})]+\gamma \oint_{C}\Omega _{2}\omega
_{21}-\bar{k}\oint_{C}K\Omega _{2}\omega _{1}.
\end{eqnarray}

Otherwise, $\omega _{13} =a\omega _{1}+b\omega
_{2}$ implies that: $ad\omega _{1}+db\wedge
\omega _{2}+2bd\omega _{2}-cd\omega
_{1}=-da\wedge \omega _{1}$. Thus
\begin{eqnarray}
&&a\Omega _{2}d\omega _{1}-bd\Omega _{2}\wedge \omega _{2}+b\Omega
_{2}d\omega _{2}+cd\Omega _{2}\wedge \omega _{1}+d(\Omega _{13}\omega
_{2}-\Omega _{23}\omega _{1}) \notag\\
&=&-d(a+c)\wedge \Omega _{2}\omega _{1} =-d(2H+c_{0})\wedge \Omega
_{2}\omega _{1}\texttt{ ($c_{0}$ is a constant)}.
\end{eqnarray}
Therefore
\begin{eqnarray}
\delta F
&=&\oint_{C}[-\frac{k_{c}}{2}(2H+c_{0})^{2}\Omega _{2}\omega _{1}-\bar{k}%
K\Omega _{2}\omega _{1}-\lambda \Omega _{2}\omega
_{1}+\gamma \Omega _{2}\omega _{21}].
\end{eqnarray}
It follows that:
\begin{equation}
\left. \lbrack \frac{k_{c}}{2}(2H+c_{0})^{2}+\bar{k}K+\lambda
+\gamma k_{g}]\right\vert _{C}=0,\label{bound3}
\end{equation}
because of $\omega_{21}=-k_g\omega_1$ on $C$.
This equation is the force equilibrium equation
of points on $C$ along the direction of
$\mathbf{e}_2$ \cite{Capovilla,Capovilla2}.

Eqs.(\ref{equlib}), (\ref{bound1}),
(\ref{bound2}) and (\ref{bound3}) are the
equilibrium equation and boundary conditions of
the membrane. They correspond to Eqs. (17), (60),
(59) and (58) in Ref.\cite{Capovilla},
respectively. In fact, these equations can be
applied to the membrane with several edges also,
because in above discussion the edge is a general
edge. But it is necessary to notice the right
direction of the edges. We call these equations
the basic equations.

\section{\label{shape}Special solutions to basic equations and their corresponding shapes}

In this section, we will give some special
solutions to the basic equations together with
their corresponding shapes. For convenience, we
consider the axial symmetric surface with axial
symmetric edges. Zhou has considered the similar
problem in his PhD thesis \cite{zhou}. If
expressing the surface in 3-dimensional space as
$\mathbf{r}=\{v\cos u,v\sin u,z(v)\}$ we obtain
$2H=-(\frac{\sin \psi }{v}+\cos \psi \frac{d\psi
}{dv})$, $K=\frac{\sin \psi \cos \psi
}{v}\frac{d\psi }{dv}$, $\nabla(2H)
=-\frac{\mathbf{r}_{2}}{\sec ^{2}\psi }\frac{d}{d
v}(\frac{\sin \psi }{v}+\cos \psi \frac{d\psi
}{dv})$ and $\nabla ^{2}(2H)=-\frac{\cos \psi
}{v}\frac{d }{d v}[v\cos \psi \frac{d }{d
v}(\frac{\sin \psi }{v}+\cos \psi \frac{d\psi
}{dv})]$, where $\psi=\arctan
[\frac{dz(v)}{dv}]$,
$\mathbf{r}_{2}=\partial\mathbf{r}/\partial v$.
Define $\mathbf{t}$ as the direction of curve $C$
and $\mathbf{r}_{1}=\partial\mathbf{r}/\partial
u$. Obviously, $\mathbf{t}$ is parallel or
antiparallel to $\mathbf{r}_{1}$ on curve $C$.
Introduce a notation $sn$, such that $sn=+1$ if
$\mathbf{t}$ is parallel to $\mathbf{r}_{1}$, and
$sn=-1$ if not. Thus $\mathbf{e}_{2}=sn
\frac{\mathbf{r}_{2}}{\sec\psi}$ and
$\mathbf{e}_{2}\cdot \nabla (2H)=-sn\cos \psi \frac{\partial }{\partial v}(\frac{\sin \psi }{v}%
+\cos \psi \frac{d\psi }{dv})$  on curve $C$. For
curve $C$, $k_n=-\frac{\sin \psi }{v}$,
$\tau_g=0$, and $k_{g}=-sn \frac{\cos \psi }{v}$.
Thus we can reduce Eqs.(\ref{equlib}),
(\ref{bound1}), (\ref{bound2}) and (\ref{bound3})
to:
\begin{eqnarray}
k_{c}(\frac{\sin \psi }{v}&+&\cos \psi \frac{d\psi
}{dv}-c_{0})[\frac{1}{2}( \frac{\sin \psi }{v}+\cos \psi
\frac{d\psi }{dv})^{2}+\frac{1}{2}c_{0}(\frac{ \sin \psi }{v}+\cos
\psi \frac{d\psi }{dv})-\frac{2\sin \psi \cos \psi }{v}
\frac{d\psi }{dv}] \nonumber\\&-&\lambda (\frac{\sin \psi
}{v}+\cos \psi \frac{d\psi }{dv})+k_{c}\frac{\cos \psi }{v}\frac{d
}{d v}[v\cos \psi \frac{d}{ d v}(\frac{\sin \psi }{v}+\cos \psi
\frac{d\psi }{dv})]=0,\label{sequilib}
\\
&&\left[k_{c}(\frac{\sin \psi }{v}+\cos \psi \frac{d\psi
}{dv}-c_{0})+\bar{k}\frac{
\sin \psi }{v}\right]_C=0,\label{sbound1}\\
&&\left[-sn \bar{k}\cos \psi \frac{d}{dv}(\frac{\sin \psi
}{v})+\gamma\frac{\sin \psi
}{v}\right]_C=0,\label{sbound2}\\
&&\left[\frac{\bar{k}^2}{2k_c}(\frac{\sin \psi
}{v})^2+\bar{k}\frac{\sin \psi \cos \psi
}{v}\frac{d\psi }{dv}+\lambda-sn\gamma \frac{\cos
\psi }{v}\right]_C=0.\label{sbound3}
\end{eqnarray}

In fact, in above four equations only three of
them are independent. We usually keep
Eqs.(\ref{sequilib}), (\ref{sbound1}) and
(\ref{sbound3}) for the axial symmetric surface.
For the general case, we conjecture that there
are also three independent equations among Eqs.
(\ref{equlib}), (\ref{bound1}), (\ref{bound2})
and (\ref{bound3}).
 Eq.(\ref{sequilib}) is the same
as the equilibrium equation of axisymmetrical
closed membranes \cite{jghu,seifert}. In Ref
\cite{seifert}, a large number of numerical
solutions to Eq.(\ref{sequilib}) as well as their
classifications are discussed.

Next, Let us consider some analytical solutions
and their corresponding shapes. We merely try to
show that these shapes exist, but not to compare
with experiments. Therefore, we only consider
analytical solutions for some specific values of
parameters.

\subsection{The constant mean curvature surface}
The constant mean curvature surfaces satisfy
Eq.(\ref{equlib}) for proper values of $k_c$,
$c_0$, $K$, and $\lambda$. But
Eqs.(\ref{bound1}), (\ref{bound2}) and
(\ref{bound3}) imply $2H+c_0=0$, $k_n=0$, and
$\bar{k}K+\gamma k_g=0$ on curve $C$ if $k_c$,
$\bar{k}$, and $\gamma$ are nonzero.

For axial symmetric surfaces, $k_n=0$ requires
$\sin \psi=0$. Therefore $K=0$ which requires
$k_g=0$. Only straight line can satisfy these
conditions. It contradicts to the fact that $C$
is a closed curve. Therefore, there is no axial
symmetric open membrane with constant mean
curvature.

\subsection{The central part of a torus}
\begin{figure}[htp!]
\includegraphics[width=7cm]{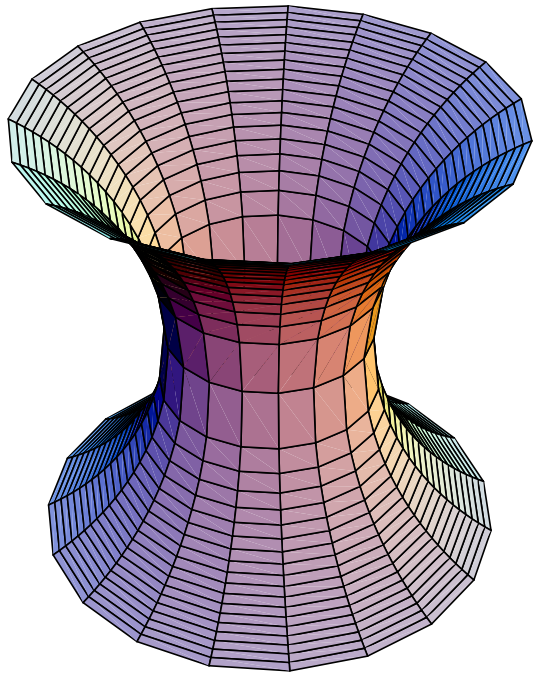}
\caption{\label{fig2}The central part of a
torus.}
\end{figure}

\begin{figure}[htp!]
\includegraphics[width=7cm]{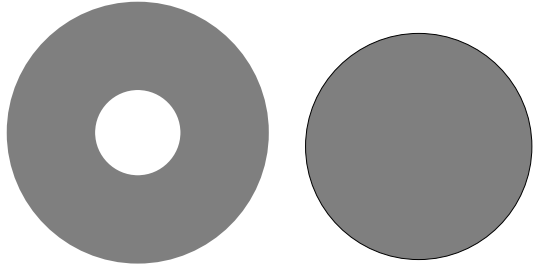}
\caption{\label{fig3}A ring (left) and a disk (right).}
\end{figure}

When $\lambda=0$, $c_0=0$, the condition
$\sin\psi=av+\sqrt{2}$ satisfies
Eq.(\ref{sequilib}). It corresponds to a torus
\cite{oy4}. Eqs.(\ref{sbound1}) and
(\ref{sbound3}) determine the position of the
edge
$v_e=-\frac{\sqrt{2}(k_c+\bar{k})}{a(2k_c+\bar{k})}$,
where
$a=-\frac{\gamma(k_c+\bar{k})\sqrt{2(2k_c^2+4k_c\bar{k}+\bar{k}^2)}}{(2k_c+\bar{k})k_c\bar{k}}$.
If we let $k_c=\bar{k}$ and
$\frac{k_c}{\gamma}=\frac{2\sqrt{14}}{3} \texttt{
(unit: length dimension)}$, it leads to $1/a=-1$
and $v_e=\frac{2\sqrt{2}}{3} \texttt{ (unit:
length dimension)}$. Thus the shape is the
central part of a torus as shown in
Fig.\ref{fig2}. This shape is topologically
equivalent to a ring as shown in Fig.\ref{fig3}.
\subsection{A cup}
\begin{figure}[htp!]
\includegraphics[width=7cm]{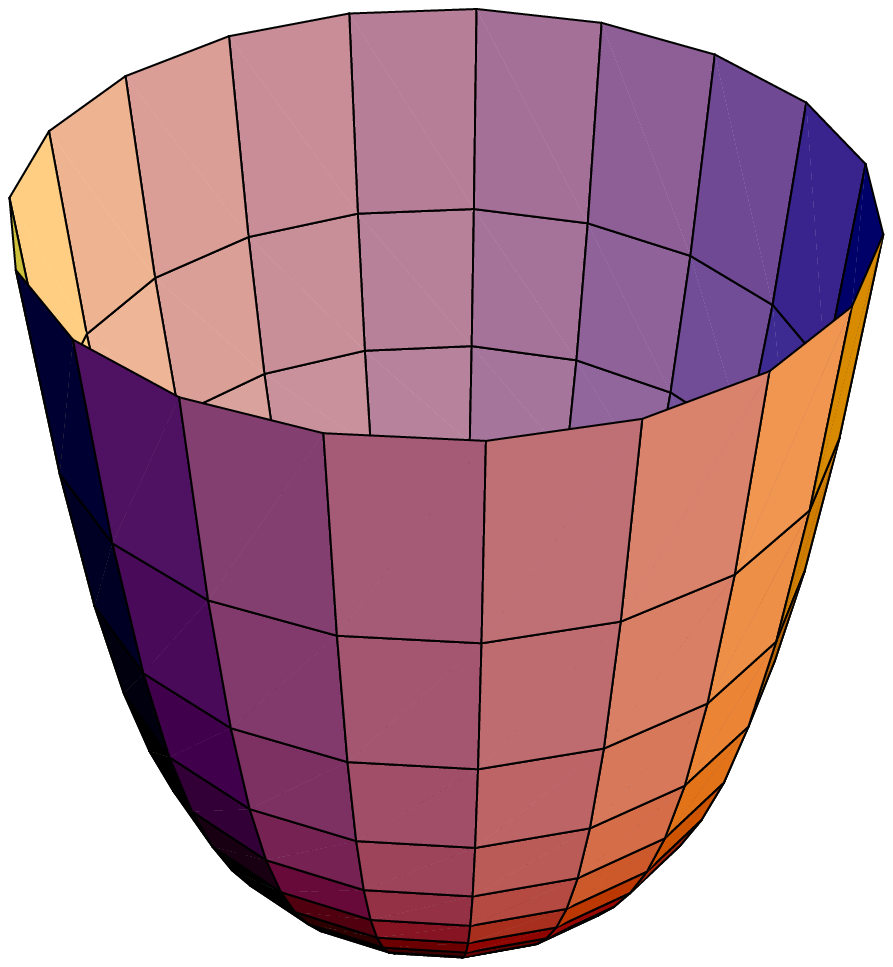}
\caption{\label{fig4}A cup.}
\end{figure}

If we let $\sin\psi=\Psi$, according Hu's method
\cite{hu} Eq.(\ref{sequilib}) reduce to:
\begin{eqnarray}
&&(\Psi ^{2}-1)\frac{d^{3}\Psi }{dv^{3}}+\Psi \frac{d^{2}\Psi }{dv^{2}}\frac{%
d\Psi }{dv}-\frac{1}{2}(\frac{d\Psi }{dv})^{3}+\frac{2(\Psi ^{2}-1)}{v}\frac{%
d^{2}\Psi }{dv^{2}}+\frac{3\Psi }{2v}(\frac{d\Psi }{dv})^{2}\nonumber \\
&&+(\frac{c_{0}^{2}}{2}+\frac{2c_{0}\Psi }{v}+\frac{\lambda }{k_{c}}-\frac{%
3\Psi ^{2}-2}{2v^{2}})\frac{d\Psi }{dv}+(\frac{c_{0}^{2}}{2}+\frac{\lambda }{%
k_{c}}-\frac{1}{v^{2}})\frac{\Psi }{v}+\frac{\Psi
^{3}}{2v^{3}}=0.\label{nequilb}
\end{eqnarray}
Now, we will consider the case that $\Psi=0$ for
$v=0$. As $v\rightarrow 0$, Eq.(\ref{nequilb})
approaches to $\frac{d^{3}\Psi
}{dv^{3}}+\frac{1}{v}\frac{d^{2}\Psi
}{dv^{2}}-\frac{1}{v^2}\frac{d\Psi
}{dv}+\frac{\Psi}{v^2}=0$. Its solution is
$\Psi=\alpha_1/v+\alpha_2v+\alpha_3v\ln v$ where
$\alpha_1=0$, $\alpha_2$ and $\alpha_3$ are three
constants. If $\lambda=0$ and $c_0>0$, we find
that $\Psi=\sin\psi=\beta (v/v0)+c_0v\ln (v/v_0)$
satisfies Eq.(\ref{sequilib}). The shapes of
closed membranes corresponding to this solution
are fully discussed by Liu \emph{et
al.}\cite{liuqh}. Eqs.(\ref{sbound1}) and
(\ref{sbound3}) determine the position of the
edge that satisfies
$\tan\psi(v)=-\frac{\gamma}{2k_cc0}$ if
$\bar{k}=-2k_c$. If let $\beta=1$,
$v_0=1/c_0=1\texttt{ (unit: length dimension)}$
and $\gamma\gg k_cc_0$, we obtain $v/v_0\approx
1$ and its corresponding shape likes a cup as
shown in Fig.\ref{fig4}. This shape is
topologically equivalent to a disk as shown in
Fig.\ref{fig3}.

\section{\label{numeric}axisymmetrical Numerical  solutions}
It is extremely difficult to find analytical
solutions to Eq.(\ref{sequilib}). We attempt to
find the numerical solutions in this section. But
there is a difficulty that $\sin\psi(v)$ is
multi-valued. To overcome this obstacle, we use
the arc-length as the parameter and express the
surface as $\mathbf{r}=\{v(s)\cos u,v(s)\sin
u,z(s)\}$. The geometrical constraint and
Eqs.(\ref{equlib}), (\ref{bound1}) and
(\ref{bound3}) now become:
\begin{eqnarray}
&&v'(s)=\cos\psi(s),\quad z'(s)=\sin\psi(s),\label{neqb1}\\
&&(2-3\sin ^{2}\psi )\psi ^{\prime }v-\sin \psi (1+\cos ^{2}\psi
)+[(c_{0}^{2}+2\lambda/k_c )\psi ^{\prime }-(\psi ^{\prime
})^{3}-2\psi ^{\prime
\prime \prime }]v^{3}\nonumber \\
&&\qquad+[(c_{0}^{2}+2\lambda/k_c )\sin \psi -4c_{0}\sin \psi \psi
^{\prime }+3\sin \psi (\psi ^{\prime })^{2}-4\cos \psi \psi
^{\prime \prime }]v^{2}=0,\label{neqb2}\\
&&\left[ k_{c}(c_{0}-\frac{\sin \psi }{v}-\psi ^{\prime
})-\bar{k}\frac{\sin
\psi }{v}\right] _{C} =0,\label{nbd1} \\
&&\left[\bar{k}c_0\frac{\sin\psi}{v}-\bar{k}(1+\frac{\bar{k}}{2k_c})\frac{\sin^2\psi}{v^2}
+\lambda -sn \gamma \frac{\cos \psi }{v}\right] _{C}
=0.\label{nbd2}
\end{eqnarray}

We can numerically solve Eqs.(\ref{neqb1}) and
(\ref{neqb2}) with initial conditions $v(0)=0$,
$\psi(0)=0$, $\psi'(0)=\alpha$ and $\psi''(0)=0$
and then find the edge position through
Eqs.(\ref{nbd1}) and (\ref{nbd2}). The shape
corresponding to the solution is topologically
equivalent to a disk as shown in Fig.\ref{fig3}.
In fact, Eq.(\ref{neqb2}) can be reduce to a
second order differential equation
\cite{Capovilla,Capovilla2,zheng}, but we still
use the third order differential equation
(\ref{neqb2}) in our numerical scheme.

\begin{figure}[htp!]
\includegraphics[width=7cm]{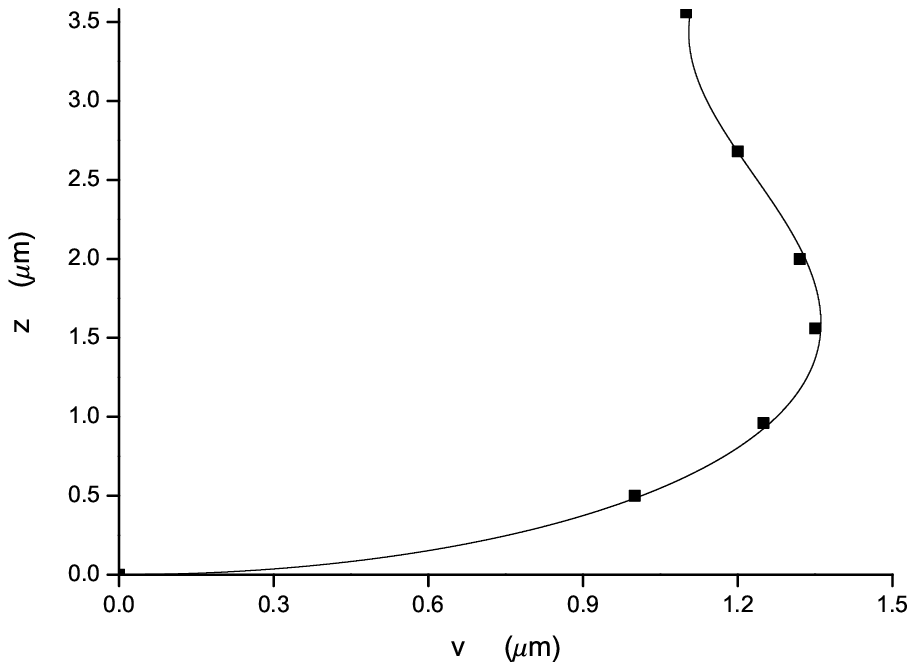}
\caption{\label{fig5}The outline of the cup-like
membrane with a wide orifice. The solid line is
the numerical result with parameters
$\alpha=c_0=0.8\mu m^{-1}$, $\lambda/k_c=0.08\mu
m^{-2}$, $\gamma/k_c=0.20\mu m^{-1}$ and
$\bar{k}/k_c=0.38$. The squares come from Fig.1d
of Ref.\cite{Hotani}. $z$-axis is the revolving
axis and $v$ is the revolving radius.}
\end{figure}

In Fig.\ref{fig5}, we depicts the outline of the
cup-like membrane with a wide orifice. The solid
line corresponds to the numerical result with
parameters $\alpha=c_0=0.8\mu m^{-1}$,
$\lambda/k_c=0.08\mu m^{-2}$, $\gamma/k_c=0.20\mu
m^{-1}$ and $\bar{k}/k_c=0.38$. The squares come
from Fig.1d of Ref.\cite{Hotani}.

\begin{figure}[htp!]
\includegraphics[width=7cm]{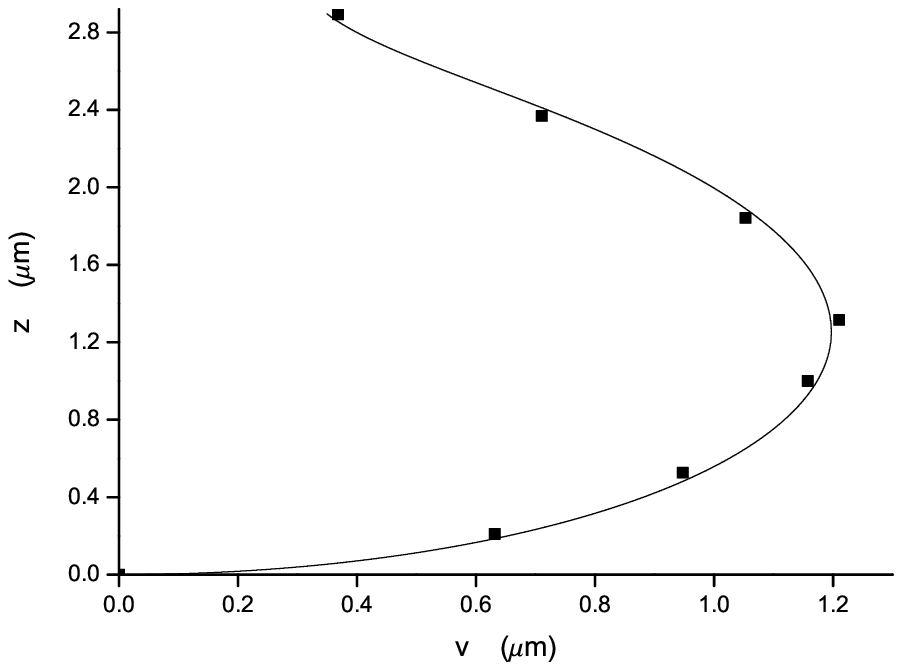}
\caption{\label{fig6}The outline of the cup-like
membrane with a narrow orifice. The solid line is
the numerical result with parameters
$\alpha=c_0=0.86\mu m^{-1}$, $\lambda/k_c=0.26\mu
m^{-2}$, $\gamma/k_c=0.36\mu m^{-1}$ and
$\bar{k}/k_c=-0.033$. The squares come from
Fig.3k of Ref.\cite{Hotani}. $z$-axis is the
revolving axis and $v$ is the revolving radius.}
\end{figure}

In Fig.\ref{fig6}, we depicts the outline of the
cup-like membrane with a narrow orifice. The
solid line corresponds to the numerical result
with parameters $\alpha=c_0=0.86\mu m^{-1}$,
$\lambda/k_c=0.26\mu m^{-2}$, $\gamma/k_c=0.36\mu
m^{-1}$ and $\bar{k}/k_c=-0.033$. The squares
come from Fig.3k of Ref.\cite{Hotani}.

Obviously, the numerical results agree quite well
with the experimental results of
Ref.\cite{Hotani}.
\section{\label{conclusion}conclusion}
In above discussion, we introduce exterior
differential forms to describe a lipid membrane
with freely exposed edge. The total free energy
is defined as the Helfrich's free energy plus the
surface and line tension energy. The equilibrium
equation and boundary conditions of the membrane
are derived from the variation of the total free
energy. These equations can also be applied to
the membrane with several freely exposed edges. A
numerical scheme to give some axisymmetric
solutions and their corresponding shapes do agree
with some experimental results.

The method that combines exterior differential
forms with the variation of surface is of
important mathematical significance. It is easy
to be generalized to deal with and to simplify
the difficult variational problems on
high-dimensional manifolds.

Although we give some axisymmetric numerical
solutions that agree with experimental results
obtained by Saitoh \emph{et al}, up to now, we
still cannot find any unsymmetrical solution. A
large number of unsymmetrical shapes are found in
experiments, which will be a challenge to the
theoretical study.
\section{\label{ackn}ACKNOWLEDGMENTS}
We are grateful to the instructive advice of Mr.
J. J. Zhou, H. J. Zhou, R. Capovilla and J.
Guven. We thank Prof. Y. Z. Xie and Dr. L. Q. Ge
for their critical reading our manuscript.

\newpage

\begin{thebibliography}{}
\bibitem{oy1}Z. C. Ou-Yang and W. Helfrich, Phys. Rev. Lett. \textbf{59},
2486 (1987); \textbf{60}, 1209 (1987); Phys. Rev. A \textbf{39}
5280 (1989).
\bibitem{oy2}H. Naito, M. Okuda, and Z. C. Ou-Yang, Phys. Rev. E \textbf{48}, 2304
(1993).
\bibitem{oy3}Z. C. Ou-Yang, Phys. Rev. A \textbf{41}, 4517 (1990).
\bibitem{Mutz} M. Mutz and D. Bensimon, Phys. Rev. A \textbf{43}, 4525
(1991); B. Fourcade, M. Mutz, and D. Bensimon Phys. Rev. Lett.
\textbf{68}, 2551 (1992).
\bibitem{Boal}D. H. Boal and M. Rao, Phys. Rev. A \textbf{46}, 3037 (1992);
D. H. Boal, \emph{Mechanics of the Cell} (Cambridge University
Press, Cambridge, England, 2002).
\bibitem{Hotani}A. Saitoh, K. Takiguchi, Y. Tanaka, and H. Hotani,
Proc. Natl. Acad. Sci. \textbf{95}, 1026 (1998).
\bibitem{Hotani2} F. Nomura, M.
Nagata, T. Inaba, H. Hiramatsu, H. Hotani, and K. Takiguchi, Proc.
Natl. Acad. Sci. \textbf{98}, 2340 (2001).
\bibitem{Capovilla}R. Capovilla, J. Guven, and J. A. Santiago, Phys. Rev. E \textbf{66}, 021607
(2002).
\bibitem{Capovilla2}R. Capovilla and J. Guven, J. Phys. A: Math. Gen.
\textbf{35}, 6233 (2002).
\bibitem{Lipowsky}F. J\"{u}licher and R. Lipowsky, Phys. Rev. Lett. \textbf{70}, 2964
(1993); F. J\"{u}licher and R. Lipowsky, Phys. Rev. E
\textbf{53}, 2670 (1996).
\bibitem{chen}S. S. Chern and W. H. Chern, \textit{Lecture on Differential
Geometry} (Peking University Press, Beijing, 1983).
\bibitem{arnold} V. I. Arnold, \textit{Mathematical Methods of Classical
Mechanics}, translated by K. Vogtmann and A. Weinstein
(Springer-Verlag, New York, 1978).
\bibitem{westenholz}C. V. Westenholz, \textit{Differential Forms in
Mathematical Physics} (North-Holland, Amsterdam, 1981).
\bibitem{helfrich}W. Helfrich, Z. Naturforsch. \textbf{28C}, 693 (1973).
\bibitem{oy4}Z. C. Ou-Yang, J. X. Liu and Y. Z. Xie, \textit{Geometric Methods in the Elastic Theory of Membranes
in Liquid Cristal Phases} (World Scientific, Singapore,
1999).
\bibitem{zhou}J. J. Zhou, PhD thesis.
\bibitem{jghu}J. G. Hu and Z. C. Ou-Yang, Phys. Rev. E \textbf{47}, 461
(1993).
\bibitem{seifert}F. J\"{u}licher and U. Seifert, Phys. Rev. E
\textbf{49}, 4728 (1994); U. Seifert, Phys. Rev. Lett.
\textbf{66}, 2404 (1991); U. Seifert, K. Berndl and R.
Lipowsky, Phys. Rev. A \textbf{44}, 1182 (1991).
\bibitem{hu}Z. N. Hu, Mod. Phys. Lett. B, \textbf{13}, 13 (1999).
\bibitem{liuqh} Q. H. Liu, H. J. Zhou, J. X. Liu, and Z. C.
Ou-Yang, Phys. Rev. E \textbf{60}, 3227 (1999).
\bibitem{zheng}W. M. Zheng and J. X. Liu, Phys. Rev. E \textbf{48}, 2856 (1993).
\end{thebibliography}
\end{document}